\documentclass[prd,superscriptaddress,amsfonts,amssymb,amsmath,showpacs,onecolumn]{revtex4-2}
\usepackage{bm}
\usepackage{amsfonts}
\usepackage{latexsym}
\usepackage{graphicx}
\usepackage{amsmath}
\usepackage{palatino}
\usepackage{mathpazo}
\usepackage{textcomp}
\usepackage{float}
\usepackage{booktabs}
\usepackage{dcolumn}
\usepackage{booktabs}
\usepackage{multirow}
\usepackage{hyperref}
\hypersetup{colorlinks,citecolor=blue}
\usepackage{amsmath}
\usepackage{xcolor}
\usepackage{orcidlink}
\usepackage{epsfig}
\usepackage{caption}
\usepackage{subcaption}
\usepackage{commath}
\def\jnl@style{\it}
\def\aaref@jnl#1{{\jnl@style#1}}

\def\aaref@jnl#1{{\jnl@style#1}}

\def\aj{\aaref@jnl{AJ}}                   
\def\apj{\aaref@jnl{ApJ}}                 
\def\apjl{\aaref@jnl{ApJ}}                
\def\apjs{\aaref@jnl{ApJS}}               
\def\apss{\aaref@jnl{Ap\&SS}}             
\def\aap{\aaref@jnl{A\&A}}                
\def\aapr{\aaref@jnl{A\&A~Rev.}}          
\def\aaps{\aaref@jnl{A\&AS}}              
\def\mnras{\aaref@jnl{Mon.~Not.~Roy.~Astron.~Soc.}}             
\def\prd{\aaref@jnl{Phys.~Rev.~D}}        
\def\prc{\aaref@jnl{Phys.~Rev.~C}}  
\def\prl{\aaref@jnl{Phys.~Rev.~Lett.}}    
\def\qjras{\aaref@jnl{QJRAS}}             
\def\skytel{\aaref@jnl{S\&T}}             
\def\ssr{\aaref@jnl{Space~Sci.~Rev.}}     
\def\zap{\aaref@jnl{ZAp}}                 
\def\nat{\aaref@jnl{Nature}}              
\def\aplett{\aaref@jnl{Astrophys.~Lett.}} 
\def\apspr{\aaref@jnl{Astrophys.~Space~Phys.~Res.}} 
\def\physrep{\aaref@jnl{Phys.~Rep.}}      
\def\physscr{\aaref@jnl{Phys.~Scr}}       
\def\commat{\aaref@jnl{Comm.~Math.~Phys.}}              
\def\science{\aaref@jnl{Science}}               
\def\cqg{\aaref@jnl{Classical Quant.~Grav.}}            
\def\jpcs{\aaref@jnl{JPCS}}                                     
\def\ijmpd{\aaref@jnl{Int.~J.~Mod.~Phys.~D}}                    
\def\grg{\aaref@jnl{Gen.~Relat.~Gravit.}}               
\def\rpp{\aaref@jnl{Rep.~Prog.~Phys.}}          
\def\npa{\aaref@jnl{Nucl.~Phys.~A}}        
\def\lrr{\aaref@jnl{Living Rev.~Rel.}}                   
\def\jcap{\aaref@jnl{J.~Cosmology Astropart.~Phys.}}    
\def\rmp{\aaref@jnl{Rev.~Mod.~Phys.}}   
\def\epjc{\aaref@jnl{Eur.~Phys.~J.~C}}

\allowdisplaybreaks[1]

\begin{document}

\color{black}       

\title{Energy conditions in $f(Q, L_m)$ gravity}

\author{Y. Myrzakulov\orcidlink{0000-0003-0160-0422}}\email[Email: ]{ymyrzakulov@gmail.com} 
\affiliation{Department of General \& Theoretical Physics, L.N. Gumilyov Eurasian National University, Astana, 010008, Kazakhstan.}

\author{O. Donmez\orcidlink{0000-0001-9017-2452}}
\email[Email: ]{orhan.donmez@aum.edu.kw}
\affiliation{College of Engineering and Technology, American University of the Middle East, Egaila 54200, Kuwait.}

\author{M. Koussour\orcidlink{0000-0002-4188-0572}}
\email[Email: ]{pr.mouhssine@gmail.com}
\affiliation{Department of Physics, University of Hassan II Casablanca, Morocco.}

\author{S. Muminov\orcidlink{0000-0003-2471-4836}}
\email[Email: ]{sokhibjan.muminov@gmail.com}
\affiliation{Mamun University, Bolkhovuz Street 2, Khiva 220900, Uzbekistan.}

\author{D. Ostemir}\email[Email: ]{ostemirdauren02@gmail.com} 
\affiliation{Department of General \& Theoretical Physics, L.N. Gumilyov Eurasian National University, Astana, 010008, Kazakhstan.}

\author{J. Rayimbaev\orcidlink{0000-0001-9293-1838}}
\email[Email: ]{javlon@astrin.uz}
\affiliation{New Uzbekistan University, Movarounnahr Street 1, Tashkent 100007, Uzbekistan.}
\affiliation{Urgench State University, Kh. Alimjan Str. 14, Urgench 221100, Uzbekistan.}

\begin{abstract}
We are experiencing a golden age of experimental cosmology, with exact and accurate observations being used to constrain various gravitational theories like never before. Alongside these advancements, energy conditions play a crucial theoretical role in evaluating and refining new proposals in gravitational physics. We investigate the energy conditions (WEC, NEC, DEC, and SEC) for two $f(Q, L_m)$ gravity models using the FLRW metric in a flat geometry. Model 1, $f(Q, L_m) = -\alpha Q + 2L_m + \beta$, features linear parameter dependence, satisfying most energy conditions while selectively violating the SEC to explain cosmic acceleration. The EoS parameter transitions between quintessence, a cosmological constant, and phantom energy, depending on $\alpha$ and $\beta$. Model 2, $f(Q, L_m) = -\alpha Q + \lambda (2L_m)^2 + \beta$, introduces nonlinearities, ensuring stronger SEC violations and capturing complex dynamics like dark energy transitions. While Model 1 excels in simplicity, Model 2's robustness makes it ideal for accelerated expansion scenarios, highlighting the potential of $f(Q, L_m)$ gravity in explaining cosmic phenomena.

\textbf{Keywords: }$f(Q,L_m)$ gravity, energy conditions, EoS parameter, dark energy.
\end{abstract}

\maketitle


\section{Introduction}\label{sec1}

Recent advancements in cosmology have highlighted the accelerating expansion of the universe \cite{Riess/1998,Riess/2004,Perlmutter/1999,T.Koivisto,S.F.,Spergel,R.R.,Z.Y.,D.J.,W.J.}, a phenomenon extensively studied and attributed to the influence of dark energy (DE). DE, characterized by a significant negative pressure and accounting for roughly 70\% of the total energy and matter content of the universe, influences the dynamics of ordinary matter through its exotic properties. This mysterious force violates the strong energy condition (SEC), expressed as $\rho + 3p \geq 0$, where $\rho$ is the energy density and $p$ is the pressure. The simplest explanation for DE is the introduction of a cosmological constant (CC), which accounts for the observed accelerated expansion and underpins the successful $\Lambda$CDM model \cite{Zlatev/1999}. However, the model faces challenges, including a mismatch between predicted and observed CC values and the cosmic coincidence problem, which questions the nearly equal densities of matter and DE in the current universe. These issues motivate the exploration of alternative models to better explain the universe's accelerated expansion \cite{Weinberg/1989,Padmanabhan/2003,Steinhardt/1999}. Theoretical approaches to explaining this expansion generally fall into two categories: (1) introducing new matter components with a negative equation of state parameter (e.g., quintessence, phantom fields, etc.) on the right-hand side of Einstein's equations, or (2) modifying the left-hand side by altering the Einstein-Hilbert action using an arbitrary function $f$, reflecting a purely geometric nature. Examples of such approaches include $f(R)$ gravity \cite{Buchdahl/1970,Dunsby/2010,Carroll/2004}, $f(R, L_m)$ gravity \cite{Harko/2010,Wang/2012,Goncalves/2023,Myrzakulova/2024,Myrzakulov/2024}, $f(G)$ gravity \cite{Felice/2009, Bamba/2017, Goheer/2009}, $f(R, \mathcal{T})$ gravity \cite{Harko/2011, Koussour_1/2022, Koussour_2/2022, Myrzakulov/2023, KK1}, $f(Q)$ gravity \cite{Jimenez/2018,Jimenez/2020,Khyllep/2021,MK1, MK2, MK3, MK4, MK5, MK6, MK7, MK8}, and $f(Q,\mathcal{T})$ gravity \cite{Xu/2019,Xu/2020,K6,K7,Bourakadi,KK2,KK3}, among others.

The $f(Q, L_m)$ gravity represents a modern extension of symmetric teleparallel gravity, where the Lagrangian density is expressed as an arbitrary function of the non-metricity $Q$ and the matter Lagrangian $L_m$ \cite{Hazarika/2024}. This coupling leads to the non-conservation of the energy-momentum tensor, resulting in significant thermodynamic implications for the universe, similar to the effects observed in $f(R, \mathcal{T})$ gravity \cite{Harko/2011}. Specific cosmological models have been investigated for various functional forms of $f(Q, L_m)$, such as $f(Q, L_m) = -\alpha Q + 2L_m + \beta$ and $f(Q, L_m) = -\alpha Q + (2L_m)^2 + \beta$ \cite{Hazarika/2024}. Comparative analyses with the standard $\Lambda$CDM paradigm have been carried out to evaluate their observational consequences. These models, developed within a flat FLRW spacetime, successfully account for the observed accelerating expansion of the universe, providing a compelling alternative explanation for the phenomenon of cosmic acceleration \cite{Hazarika/2024}. The inclusion of $L_m$ in $f(Q, L_m)$ gravity is motivated by the exploration of nonminimal couplings between geometry and matter. Such couplings have been studied in various modified gravity theories, including $f(R, L_m)$ \cite{Harko/2010} and $f(R, L_m, T)$ gravity \cite{fRLT}, as a means to investigate potential deviations from general relativity (GR) and possible explanations for DE and dark matter \cite{Mustafa}. Recently, Myrzakulov et al. \cite{fQL1} explored the universe's late-time expansion within the $f(Q, L_m)$ gravity framework. They analyzed a linear model, $f(Q, L_m) = -\alpha Q + 2 L_m + \beta$, and constrained the parameters $H_0$, $\alpha$, and $\beta$ using datasets such as $H(z)$, Pantheon+SH0ES, and BAO. Their findings indicate a positive energy density approaching zero in the distant future and a deceleration parameter that transitions from deceleration to acceleration, aligning with previous observational studies and enhancing our understanding of the universe's expansion dynamics. In Ref. \cite{fQL2}, the authors analyzed a model characterized by $f(Q, L_m) = \alpha Q + \beta L_m^n$. Their analysis reveals that the universe undergoes accelerated expansion for $n > 2$, while the case $n = 1$ corresponds to a universe dominated by non-relativistic matter. Myrzakulov et al. \cite{fQL3} investigated the impact of bulk viscosity on late-time cosmic acceleration within the $f(Q, L_m)$ gravity framework. They analyzed the function $f(Q, L_m) = \alpha Q + \beta L_m$ and derived exact solutions under non-relativistic matter domination. 

This study aims to investigate the various energy conditions within the newly formulated $f(Q, L_m)$ gravity theory. In GR, energy conditions are crucial in understanding key aspects of cosmology \cite{Visser/2000}, black hole thermodynamics \cite{Hawking/1973}, and singularity theorems \cite{Wald/1984}. These conditions represent different approaches to ensuring the positivity of the energy-momentum tensor in the presence of matter, while also capturing the attractive nature of gravity. Derived from the Raychaudhuri equation, energy conditions are fundamentally geometric, requiring that gravity is attractive and the energy density remains positive \cite{Santos/2007}. The weak, null, dominant, and strong energy conditions are fundamental in GR. The renowned Hawking-Penrose singularity theorem relies on the SEC, whose violation leads to the observed accelerated expansion of the universe \cite{Visser/2000}. The second law of black hole thermodynamics is based on the null energy condition (NEC) \cite{Wald/1984,Visser/1996}. The weak energy condition (WEC) is essentially a combination of the NEC ($\rho + p \geq 0$) and $\rho \geq 0$, meaning the local energy density, as measured by any timelike observer, must be positive. If the NEC is violated, the SEC is no longer satisfied, and likewise, the WEC and dominant energy condition (DEC) cannot hold. By focusing on a congruence of null geodesics, one can derive the null convergence condition $R_{\mu\nu} k^\mu k^\nu \geq 0$ for any null vector $k^\mu$.

Energy conditions have been extensively studied in the context of various modified gravity theories. For instance, Capozziello et al. \cite{Capozziello/2018} analyzed energy conditions in GR under the framework of $f(R)$ gravity with a power-law formulation. Atazadeh et al. \cite{Atazadeh/2009} examined the implications of energy conditions in Brans-Dicke's theory derived from a generic $f(R)$ model. Liu and Reboucas \cite{Liu/2012} explored energy conditions in $f(T)$ gravity, applying constraints from exponential and Born-Infeld $f(T)$ models. Zubair and Waheed \cite{Zubair/2015} investigated the validity of energy bounds in a modified gravity theory incorporating a non-minimal coupling between matter and the torsion scalar. Azizi and Gorjizadeh \cite{Azizi/2017} addressed energy conditions in higher-derivative torsion gravity. Studies on $f(G)$ gravity, including diverse forms, were carried out by Garcia et al. \cite{Garcia/2011} and Bamba et al. \cite{Bamba/2017}. Further, Sharif and Ikram \cite{Sharif/2016} discussed energy conditions for reconstructed $f(G, \mathcal{T})$ models within the FLRW universe, and Yousaf et al. \cite{Yousaf/2018} evaluated the viability of energy bounds in higher-order $f(R, \Box R, \mathcal{T})$ gravity through energy conditions. Recently, Mandal et al. \cite{Mandal/2020} analyzed energy conditions within $f(Q)$ gravity, whereas Arora et al. \cite{Arora/2021} investigated these conditions in extended $f(Q, \mathcal{T})$ gravity.

The structure of this study is organized as follows: In Section \ref{sec2}, we present the fundamental aspects of $f(Q, L_m)$ gravity. Section \ref{sec3} employs the Raychaudhuri equations to derive the energy conditions, incorporating the effects of nonmetricity and the matter Lagrangian. Detailed constraints on $f(Q, L_m)$ models are analyzed in Section \ref{sec4}, followed by a comparative discussion of the $f(Q, L_m)$ models. In addition, the equation of state parameter for the models under consideration is depicted. Finally, Section \ref{sec5} concludes with our remarks and future outlook.

\section{Overview of $f(Q,L_{m})$ Gravity Theory}\label{sec2}

In this work, we explore an extension of symmetric teleparallel gravity, with the action formulated as described in \cite{Hazarika/2024},
\begin{equation}
    S=\int f(Q,L_m) \sqrt{-g} d^4x. \label{Action}
\end{equation}

Here, $\sqrt{-g}$ represents the square root of the negative determinant of the metric, while $f(Q, L_m)$ is an arbitrary function of the non-metricity scalar $Q$ and the matter Lagrangian $L_m$. As discussed by Jimenez et al. \cite{Jimenez/2018}, the non-metricity function is defined as
\begin{equation}
Q\equiv -g^{\mu \nu }(L_{\,\,\,\alpha \mu }^{\beta }L_{\,\,\,\nu \beta
}^{\alpha }-L_{\,\,\,\alpha \beta }^{\beta }L_{\,\,\,\mu \nu }^{\alpha }),
\label{2}
\end{equation}%
where $L_{\alpha \gamma }^{\beta }$ is the disformation tensor, explicitly given by
\begin{equation}
L_{\alpha \gamma }^{\beta }=\frac{1}{2}g^{\beta \eta }\left( Q_{\gamma
\alpha \eta }+Q_{\alpha \eta \gamma }-Q_{\eta \alpha \gamma }\right).  \label{3}
\end{equation}

Another essential component of symmetric teleparallel gravity is the non-metricity tensor, defined as
\begin{equation}
Q_{\gamma \mu \nu }=-\nabla _{\gamma }g_{\mu \nu }=-\partial _{\gamma
}g_{\mu \nu }+g_{\nu \sigma }\widetilde{\Gamma }{^{\sigma }}_{\mu \gamma
}+g_{\sigma \mu }\widetilde{\Gamma }{^{\sigma }}_{\nu \gamma }.  \label{4}
\end{equation}%

Here, $\widetilde{\Gamma }{^{\gamma }}_{\mu \nu }$ denotes the Weyl connection, with $\Gamma {^{\gamma }}_{\mu \nu }$ representing the familiar Levi-Civita connection associated with the metric. Moreover, the trace of the non-metricity tensor is expressed as
\begin{equation}
Q_{\beta }=g^{\mu \nu }Q_{\beta \mu \nu },\qquad \widetilde{Q}_{\beta
}=g^{\mu \nu }Q_{\mu \beta \nu }.  \label{5}
\end{equation}%

Next, we define the superpotential tensor, also known as the non-metricity conjugate, as
\begin{equation}
\hspace{-0.5cm} P_{\ \ \mu \nu }^{\beta }\equiv \frac{1}{4}\bigg[-Q_{\ \
\mu \nu }^{\beta }+2Q_{\left( \mu \ \ \ \nu \right) }^{\ \ \ \beta
}+Q^{\beta }g_{\mu \nu }-\widetilde{Q}^{\beta }g_{\mu \nu }  
\hspace{-0.5cm} -\delta _{\ \ (\mu }^{\beta }Q_{\nu )}\bigg]=-\frac{1}{2}%
L_{\ \ \mu \nu }^{\beta }+\frac{1}{4}\left( Q^{\beta }-\widetilde{Q}^{\beta
}\right) g_{\mu \nu }-\frac{1}{4}\delta _{\ \ (\mu }^{\beta }Q_{\nu )}.\quad
\quad   \label{6}
\end{equation}

Further, the non-metricity scalar is defined as \cite{Jimenez/2018}
\begin{equation}
Q=-Q_{\beta \mu \nu }P^{\beta \mu \nu }=-\frac{1}{4}\big(-Q^{\beta \nu
\rho }Q_{\beta \nu \rho }+2Q^{\beta \nu \rho }Q_{\rho \beta \nu } 
-2Q^{\rho }\tilde{Q}_{\rho }+Q^{\rho }Q_{\rho }\big).  \label{7}
\end{equation}

Therefore, varying the action (\ref{Action}) with respect to the metric yields the field equations,
\begin{equation}
\frac{2}{\sqrt{-g}}\nabla_\alpha(f_Q\sqrt{-g}P^\alpha_{\;\;\mu\nu}) +f_Q(P_{\mu\alpha\beta}Q_\nu^{\;\;\alpha\beta}-2Q^{\alpha\beta}_{\;\;\;\mu}P_{\alpha\beta\nu})
+\frac{1}{2}f g_{\mu\nu}=\frac{1}{2}f_{L_m}(g_{\mu\nu}L_m-T_{\mu\nu}),\label{field}
\end{equation}
where $f_Q=\partial f(Q,L_m)/\partial Q$ and $f_{L_m}=\partial f(Q,L_m)/\partial L_m$. 

For $f(Q, L_m) = f(Q) + 2L_m$, the field equations reduce to those of $f(Q)$ gravity \cite{Jimenez/2018}. Furthermore, the energy-momentum tensor $T_{\mu\nu}$ for the matter is given by
\begin{equation}
    T_{\mu\nu}=-\frac{2}{\sqrt{-g}}\frac{\delta(\sqrt{-g}L_m)}{\delta g^{\mu\nu}}=g_{\mu\nu}L_m-2\frac{\partial L_m}{\partial g^{\mu\nu}},
\end{equation}

By varying the gravitational action with respect to the connection, we again obtain the field equations,
\begin{equation}
    \nabla_\mu\nabla_\nu\Bigl( 4\sqrt{-g}\,f_Q\,P^{\mu\nu}_{\;\;\;\;\alpha}+H_\alpha^{\;\;\mu\nu}\Bigl)=0,
\end{equation}
where $H_\alpha^{\;\;\mu\nu}$ denotes the hypermomentum density, defined as
\begin{equation}
    H_\alpha^{\;\;\mu\nu}=\sqrt{-g}f_{L_m}\frac{\delta L_m}{\delta Y^\alpha_{\;\;\mu\nu}}.
\end{equation}

By applying the covariant derivative to the field equation (\ref{field}), one can obtain
\begin{equation}
D_\mu\,T^\mu_{\;\;\nu}= \frac{1}{f_{L_m}}\Bigl[ \frac{2}{\sqrt{-g}}\nabla_\alpha\nabla_\mu H_\nu^{\;\;\alpha\mu} + \nabla_\mu\,A^{\mu}_{\;\;\nu} - \nabla_\mu \bigr( \frac{1}{\sqrt{-g}}\nabla_\alpha H_\nu^{\;\;\alpha\mu}\bigr) \Bigr]=B_\nu \neq 0.
\end{equation}

In the $f(Q, L_m)$ gravity, the matter energy-momentum tensor is not conserved. The non-conservation tensor $B_\nu$ depends on dynamical variables such as $Q$, $L_m$, and the thermodynamic quantities of the system.

To examine the energy conditions in $f(Q,L_m)$ gravity, we consider the universe within the context of flat FLRW geometry. This model is based on two key assumptions about the universe's structure \cite{ryden/2003}: (i) Homogeneity: the universe is uniform at any given moment, with its density and structure being consistent across large scales, indicating the absence of preferred locations. (ii) Isotropy: observations from any point in the universe show identical physical laws and conditions in all directions, suggesting that the universe appears the same from any viewpoint. In FLRW geometry, the spacetime interval is expressed by the line element:
\begin{equation}
\label{FLRW}
    ds^2=-dt^2+a^2(t)(dx^2+dy^2+dz^2),
\end{equation}
where $a(t)$ represents the scale factor, which evolves with cosmic time $t$. From the metric (\ref{FLRW}), the non-metricity scalar is given by $Q = 6 H^2$, where $H = \frac{\dot{a}}{a}$ is the Hubble parameter, describing the rate of the universe's expansion.

Furthermore, the matter content of the universe is assumed to be a perfect fluid (without viscosity, shear stresses, or heat flux), with its energy-momentum tensor given by
\begin{equation}
    T_{\mu\nu}=(\rho+p)u_{\mu}u_{\nu}+pg_{\mu\nu},
    \label{EMT}
\end{equation}
where $\rho$ is the energy density, $p$ is the isotropic pressure, and $u_\mu$ is the four-velocity of the fluid.

Therefore, by substituting Eqs. (\ref{FLRW}) and (\ref{EMT}), we obtain the modified Friedmann equations in $f(Q, L_m)$ gravity, which are explicitly given by \cite{Hazarika/2024,fQL1,fQL2,fQL3}
\begin{eqnarray}
\label{F1}
    && 3H^2 =\frac{1}{4f_Q}\bigr[ f - f_{L_m}(\rho + L_m) \bigl],\\
   && \dot{H} + 3H^2 + \frac{\dot{f_Q}}{f_Q}H=\frac{1}{4f_Q}\bigr[ f + f_{L_m}(p - L_m) \bigl]. \label{F2}
\end{eqnarray}

Particularly, when $f(Q, L_m) = f(Q) + 2L_m$, the Friedmann equations simplify to those of $f(Q)$, which can then be further reduced to the symmetric teleparallel version of GR. In addition, it can be shown that the density and pressure also satisfy the generalized energy balance equation for $f(Q, L_m)$ gravity \cite{Hazarika/2024}, which takes the form
\begin{equation}
\dot{\rho} + 3 H (\rho + p) = B_\mu u^\mu.    
\label{bal}
\end{equation}

It is clear that Eq. (\ref{bal}) differs significantly from the standard form, with additional terms on the right-hand side that account for deviations from geodesic motion. In this context, the source term, $B_\mu u^\mu$, is related to the generation or dissipation of energy. When $B_\mu u^\mu = 0$, the system follows the energy conservation law of standard gravity. However, energy transfer processes become dominant if $B_\mu u^\mu$ is nonzero.

Using Eq. \eqref{F1}, Eq. \eqref{F2} can be rewritten as
\begin{equation}
2\dot{H}+3H^2=\frac{1}{4f_Q}\left[f+f_{L_m}\left(\rho+2p-L_m\right)\right]-2\frac{\dot{f}_Q}{f_Q}H.
\end{equation}

Therefore, the generalized Friedmann equations of $f\left(Q, L_m\right)$ gravity can be reformulated as
\begin{equation}\label{41}
3H^2=\rho_{eff}, \ 2\dot{H}+3H^2=-p_{eff}.
\end{equation}

Here, we have introduced the effective energy density and pressure, defined as
\begin{equation}
\label{F1_eff}
\rho_{eff}=\frac{1}{4f_Q}\bigr[ f - f_{L_m}(\rho + L_m) \bigl],
\end{equation}
and
\begin{equation}
\label{F2_eff}
p_{eff}=2\frac{\dot{f}_Q}{f_Q}H-\frac{1}{4f_Q}\left[f+f_{L_m}\left(\rho+2p-L_m\right)\right],
\end{equation}
respectively. This behavior will be used in the next section to discuss the physical interpretations of the different energy conditions. Further, Eq. (\ref{41}) allows for the formulation of the generalized effective conservation equation in $f\left(Q, L_m\right)$ gravity as
\begin{equation}\label{cons1}
\dot{\rho}_{eff}+3H\left(\rho_{eff}+p_{eff}\right)=0.
\end{equation}

\section{The Raychaudhuri equation and energy conditions} \label{sec3}

Energy conditions in modified gravity serve as tools that govern spacetime's causal and geodesic structure. These conditions are derived using the Raychaudhuri equations, which describe the behavior of congruences and the attractive nature of gravity for timelike, spacelike, or lightlike geodesics. We will derive this equation in the Weyl framework to analyze the implications of non-metricity and the coupling of the matter Lagrangian $L_m$ within the Raychaudhuri equation. Weyl geometry, where the orientation and magnitude of a vector can vary under parallel transport, serves as a natural setup to explore the effects of non-metricity and the coupling to matter dynamics. Iosifidis et al. \cite{Iosifidis/2018} generalized the Raychaudhuri equation to spacetimes with torsion and nonmetricity, extending its conventional formulation in Riemannian geometry. They examined the influence of these geometric features on the equation's structure, emphasizing their effects on geodesic congruences and kinematic quantities, including expansion, shear, and vorticity. Furthermore, Yang et al. \cite{Yang/2021} investigated the geodesic deviation, Raychaudhuri equation, Newtonian limit, and tidal forces within the framework of Weyl-type $f(Q, \mathcal{T})$ gravity. They extend key geometric and physical concepts from GR to this modified gravity theory, exploring the effects of nonmetricity and matter-geometry coupling on these phenomena. Inspired by the above works, Arora et al. \cite{Arora/2021} derived the Raychaudhuri equation within the Weyl framework. In this study, we follow a similar approach to derive the Raychaudhuri equation in $f(Q, L_m)$ gravity. First, the Weyl connection is given by \cite{Yang/2021}
\begin{equation}
\widetilde{\Gamma}^{\gamma}_{\mu\nu} \equiv \Gamma^{\gamma}_{\mu\nu} + g_{\mu\nu} w^{\gamma} - \delta^{\gamma}_{\mu} w_{\nu} - \delta^{\gamma}_{\nu} w_{\mu},
\end{equation}
where $w_{\mu}$ is the Weyl vector field. In this framework, the metric tensor's covariant derivatives satisfy:
\begin{equation}
\widetilde{\nabla}_{\gamma} g_{\mu\nu} = 2 w_{\gamma} g_{\mu\nu}, \quad \widetilde{\nabla}_{\gamma} g^{\mu\nu} = -2 w_{\gamma} g^{\mu\nu}.
\end{equation}

Thus, the non-metricity tensors become:
\begin{equation}
Q_{\gamma\mu\nu} = -\widetilde{\nabla}_{\gamma} g_{\mu\nu} = -2 w_{\gamma} g_{\mu\nu}, \quad Q^{\gamma\mu\nu} = \widetilde{\nabla}^{\gamma} g^{\mu\nu} = -2 w^{\gamma} g^{\mu\nu},
\end{equation}
and the non-metricity scalar is
\begin{equation}
Q = -6 w^2.
\end{equation}

Here, $w^2 = g_{\mu\nu} w^\mu w^\nu$ is the norm of the Weyl vector field. In the presence of non-metricity, the length of vectors changes under parallel transport. The four-velocity is given by
\begin{equation}
u_{\mu} u^{\mu} = g_{\mu\nu} u^{\mu} u^{\nu} = -l^2, \quad l = l(x^{\alpha}),
\end{equation}
where $u^\mu = dx^\mu / d\lambda$, $\lambda$ is the affine parameter, and $l(x^\alpha)$ is a function of spacetime coordinates. The projection tensor becomes
\begin{equation}
h_{\mu\nu} = g_{\mu\nu} + \frac{1}{l^2} u_{\mu} u_{\nu}.
\end{equation}

The 4-acceleration and hyper 4-acceleration are defined as \cite{Iosifidis/2018}
\begin{equation}
A^\mu = u^\lambda \widetilde{\nabla}_\lambda u^\mu, \quad a_\mu = u^\lambda \widetilde{\nabla}_\lambda u_\mu,
\end{equation}
with a constraint between them:
\begin{equation}
A^\mu = a^\mu + Q^{\nu\lambda\mu} u_\nu u_\lambda.
\end{equation}

In $f(Q, L_m)$ gravity, the extra force $f^\mu$ arises from the coupling between $Q$ and $L_m$,
\begin{equation}
A^\mu = \frac{d^2 x^\mu}{d\lambda^2} + \widetilde{\Gamma}^\mu_{\nu\lambda} u^\nu u^\lambda = f^\mu.
\end{equation}

The Raychaudhuri equation in this framework is derived using the curvature tensor,
\begin{equation}
\left(\widetilde{\nabla}_\mu \widetilde{\nabla}_\nu - \widetilde{\nabla}_\nu \widetilde{\nabla}_\mu\right) u_\lambda = -\widetilde{R}_{\beta\lambda\nu\mu} u^\beta.
\end{equation}

Contracting with $g^{\lambda\nu} u^\mu$ and applying constraints for autoparallel curves ($ A^\mu = f^\mu = 0$), the Raychaudhuri equation simplifies to
\begin{equation}
\label{Ray}
\left(\theta - 2\frac{l'}{l}\right)' = -\frac{1}{3} \left(\theta - 2\frac{l'}{l}\right)^2 - R_{\mu\nu} u^\mu u^\nu - 2\left(\sigma^2 - \omega^2\right),
\end{equation}
where $\theta$, $\sigma^2$, and $\omega^2$ are the expansion, shear, and rotation scalars, respectively.

Following the approach introduced by Iosifidis et al.\cite{Iosifidis/2018}, we assume that in an irrotational and shear-free scenario, Eq. (\ref{Ray}) leads to the constraint
\begin{equation}
\left(\theta - 2\,\frac{l'}{l}\right)' + \frac{1}{3}\,\left(\theta - 2\,\frac{l'}{l}\right)^{2} \leq 0, \quad \text{if } R_{\mu \nu} u^{\mu} u^{\nu} \geq 0.
\end{equation}

This results in an attractive nature for gravity, providing a generalized framework for the constraints on energy conditions. Notably, by substituting $\theta-2\,\frac{l^{\prime}}{l} \rightarrow \theta$ into \eqref{Ray}, the standard Raychaudhuri equation of GR is recovered. Furthermore, the same procedure outlined here can be extended to the case of a null vector $k^{\mu}$, yielding the following simplified form of the Raychaudhuri equation,
\begin{equation}
\left(\theta-2\,\frac{l^{\prime}}{l}\right)^{\prime} = -\frac{1}{3}\,\left(\theta-2\,\frac{l^{\prime}}{l}\right)^{2}
-R_{\mu\nu}k^{\mu}k^{\nu}-2\,\left(\sigma^2-\omega^2\right).
\end{equation}

Thus, we can derive the constraint, extending the approach of Santos et al. \cite{Santos/2007}
\begin{equation}
\left(\theta-2\,\frac{l^{\prime}}{l}\right)^{\prime}+\frac{1}{3}\,\left(\theta-2\,\frac{l^{\prime}}{l}\right)^{2}\leq 0, \quad \text{if } R_{\mu \nu} k^{\mu} k^{\nu} \geqslant 0.
\end{equation}

In $f(Q, L_m)$ gravity, the effective energy density and pressure are influenced by the non-minimal coupling between $Q$ and $L_m$. The generalized energy conditions are \cite{Mandal/2020,Arora/2021}:
\begin{itemize}
    \item Weak energy condition (WEC): $ \rho_{eff} \geq 0$,
    \item Null energy condition (NEC): $ \rho_{eff} + p_{eff} \geq 0$,
    \item Dominant energy condition (DEC): $\rho_{eff} \geq |p_{eff}|$,
    \item Strong energy condition (SEC): $\rho_{eff} + 3p_{eff} \geq 0$.
\end{itemize}

By substituting Eqs. (\ref{F1_eff}) and (\ref{F2_eff}) into the previous expressions, we derive the following set of energy conditions:
\begin{itemize}
    \item Weak energy condition (WEC) $\Leftrightarrow \rho \geq 0$,
    \item Null energy condition (NEC) $\Leftrightarrow \rho + p \geq 0$,
    \item Dominant energy condition (DEC) $\Leftrightarrow \rho - p \geq 0$,
    \item Strong energy condition (SEC) $\Leftrightarrow \rho + 3p \geq 0$.
\end{itemize}

The WEC ($\rho \geq 0$) ensures that the energy density, as measured by any timelike observer, is always non-negative. In cosmological terms, this implies that the universe contains physical matter and energy with positive mass density, aligning with classical expectations. The WEC is a minimal requirement for any realistic cosmological model, ensuring that the energy density observed locally does not violate basic physical principles \cite{Santos/2007}. Its satisfaction in most standard cosmological scenarios is a cornerstone for describing ordinary matter and radiation. Further, the NEC ($\rho + p \geq 0$) ensures that the energy density and pressure along any null (lightlike) direction contribute positively. In the context of cosmology, the NEC governs the causal structure of spacetime and is fundamental to the behavior of light cones and the propagation of information. It also plays a key role in the validity of the Raychaudhuri equation, which describes the focus of geodesics. The NEC is often considered a baseline condition for avoiding exotic phenomena like closed timelike curves or violations of causality. In expanding cosmological models, the NEC dictates that energy density diminishes as the universe evolves. The DEC ($\rho - p \geq 0$) requires that the energy density exceeds or equals the pressure in magnitude. This ensures that the energy flux respects causality, meaning that energy and matter propagate at speeds less than or equal to the speed of light. Cosmologically, the DEC restricts the types of matter and energy that can exist in the universe to those consistent with causal evolution. For example, ordinary matter and radiation satisfy the DEC, while certain exotic fields, such as phantom energy, may violate it. The DEC also implies that the pressure cannot be so large as to dominate over the energy density. Finally, the SEC ($\rho + 3p \geq 0$) has a deeper connection to the dynamics of spacetime curvature and the expansion or contraction of the universe. It requires that matter and energy contribute to the focus of geodesics, a key aspect of gravitational attraction. In cosmology, the SEC is critical for understanding deceleration in an expanding universe. When the SEC holds, the combined effects of energy density and pressure work to slow cosmic expansion. However, the observed acceleration of the universe’s expansion—driven by dark energy or a cosmological constant—requires the violation of the SEC \cite{Visser/2000}. This violation is a hallmark of models that include negative-pressure components, such as quintessence or other dark energy candidates.

\section{$f(Q,L_m)$ Cosmological models}
\label{sec4}

In this section, we delve into the application of energy condition constraints to refine and restrict specific models within the framework of $f(Q,L_m)$ gravity. To effectively analyze these conditions, one critical cosmological parameter comes into play: the deceleration parameter. This parameter plays a pivotal role in characterizing the dynamics of the universe's expansion and explains its periods of acceleration or deceleration. Its mathematical definition is given as \cite{Mandal/2020,Arora/2021}
\begin{equation}
q = -1 - \frac{\dot{H}}{H^2}.    
\end{equation}
where $H$ is the Hubble parameter and $\dot{H}$ represents its time derivative. This formulation relates the deceleration parameter directly to the rate of change of the Hubble parameter: 
\begin{itemize}
    \item Decelerating phase ($q > 0$): This occurs when $\dot{H} < -H^2$, typically during matter or radiation-dominated epochs, where the expansion rate slows down over time.
    \item Accelerating phase ($q < 0$): This occurs when $\dot{H} > -H^2$, as observed in the present epoch, driven by dark energy or modifications to gravity, where the expansion accelerates. 
\end{itemize}

Further, the time derivative of the Hubble parameter can be expressed as
\begin{equation}
\label{dH}
\dot{H} = -H^2 (1 + q).    
\end{equation}

To constrain the energy conditions using phenomenological observations, we adopt the present values for the Hubble parameter and the deceleration parameter from recent Planck 2018 observations, given as $H_0=H(z=0) = 67.4$ $\text{km s}^{-1} \text{Mpc}^{-1}$ and $q_0 =q(z=0) = -0.53$, respectively \cite{Planck/2014,Planck/2020}.

To examine the nature of the matter or energy content in the universe, we introduce the EoS (equation of state) parameter, which describes the relationship between pressure $p$ and energy density $\rho$ (i.e., $\omega=\frac{p}{\rho}$) \cite{w1,w2}. The EoS plays a key role in determining the evolution of the universe and the behavior of different cosmic components. Various phases are associated with specific values of $\omega$: The dust phase occurs when $\omega = 0$, representing non-relativistic matter such as dark matter and ordinary matter, where pressure is negligible compared to energy density, and it dominates during the matter-dominated era. When $\omega = \frac{1}{3}$, the universe is in the radiation-dominated phase, where the pressure is one-third of the energy density, typical of the early universe when radiation and relativistic particles were dominant. At $\omega = -1$, we have vacuum energy, associated with the cosmological constant and dark energy in the $\Lambda$CDM model, which drives the accelerated expansion of the universe. The accelerating universe phase, where $\omega < -\frac{1}{3}$, includes the quintessence regime ($-1 < \omega < -\frac{1}{3}$) characterized by evolving dark energy, and the phantom energy regime ($\omega < -1$), which suggests an even faster cosmic acceleration and the potential for a big rip, where all structures in the universe are eventually torn apart.

\subsection{Model 1: $f(Q,L_m)=-\alpha Q+ 2L_{m}+\beta$ with $L_{m}=\rho$}

As an initial model, we consider $f(Q,L_m)=-\alpha\, Q+ 2L_{m}+\beta$, where $\alpha$ and $\beta$ are free parameters. This model, first proposed by Hazarika et al. \cite{Hazarika/2024}, naturally accounts for an accelerating, expanding universe \cite{fQL1}. For this formulation, we obtain $f_Q =- \alpha$ and $f_{L_m} = 2$. Hence, the Friedmann equations (\ref{F1}) and (\ref{F2}) simplify to
\begin{align}
\label{F1M1}
\rho &=3 \alpha H^2+\frac{\beta}{2}, \\ 
p &= - 3 \alpha H^2- 2 \alpha \dot{H}-\frac{\beta}{2}.
\label{F2M1}
\end{align}

By substituting Eqs. (\ref{F1M1}) and (\ref{F2M1}) along with Eq. (\ref{dH}) into the energy conditions, we obtain the following constraints:
\begin{equation} \label{WEC1}
WEC \Leftrightarrow 3 \alpha H_0^2+\frac{\beta}{2} \geq 0,
\end{equation}

\begin{equation} \label{NEC1}
NEC \Leftrightarrow 2 \alpha H_0^2 (q_0 + 1) \geq 0,
\end{equation}

\begin{equation} \label{DEC1}
DEC \Leftrightarrow \beta - 2 \alpha H_0^2 (q_0 - 2) \geq 0,
\end{equation}

\begin{equation}  \label{SEC1}
SEC \Leftrightarrow 6 \alpha H_0^2 q_0 - \beta \geq 0.
\end{equation}

\begin{figure}[H]
\centering
\includegraphics[width=.3\textwidth]{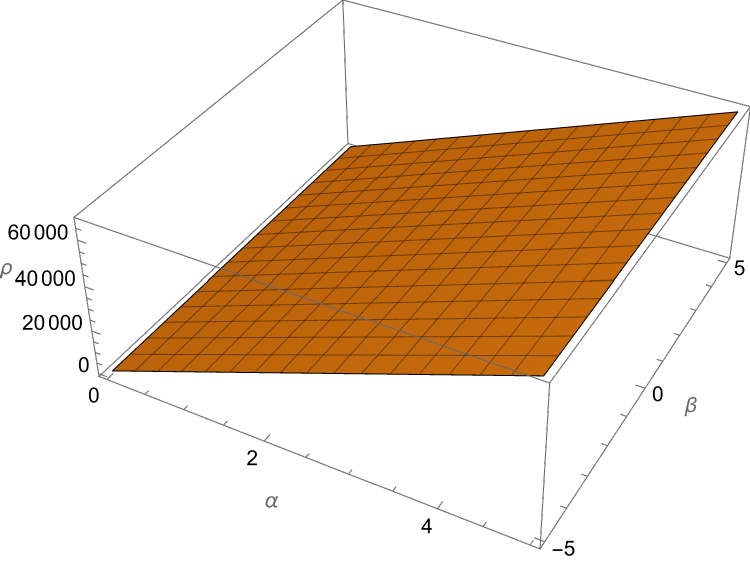}\quad
\includegraphics[width=.3\textwidth]{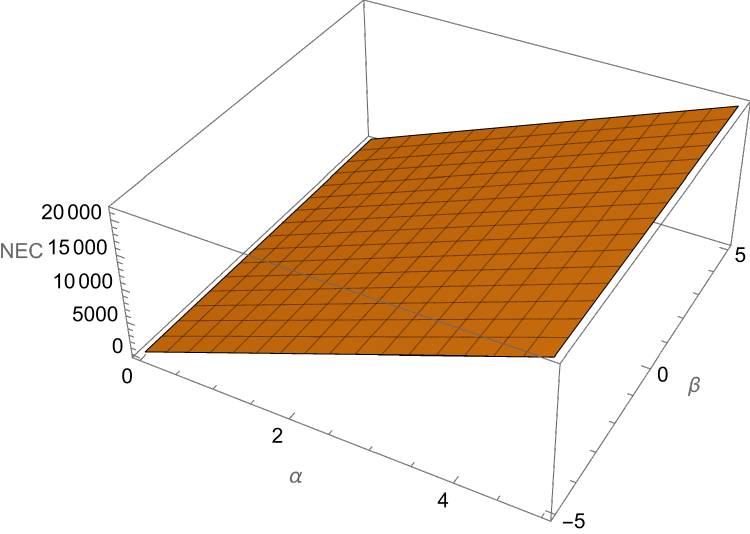}\quad
\includegraphics[width=.3\textwidth]{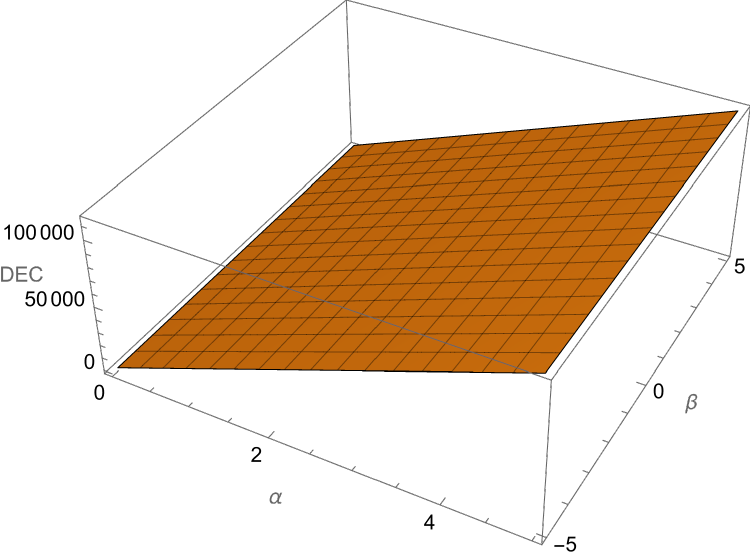}

\medskip

\includegraphics[width=.3\textwidth]{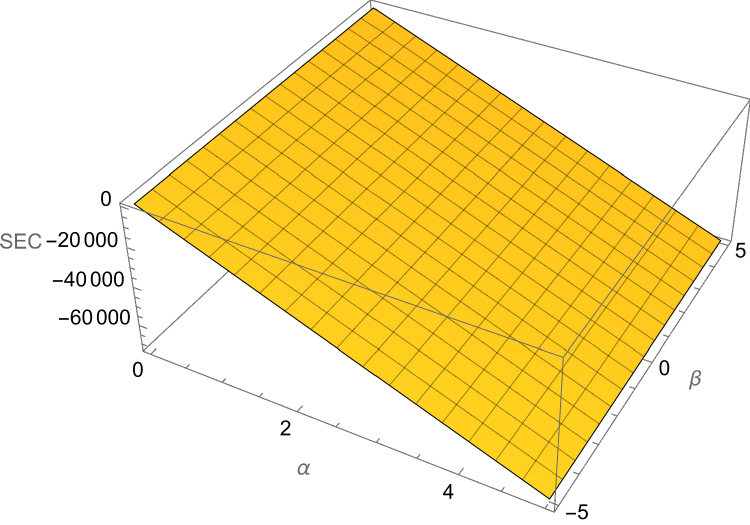}\quad
\includegraphics[width=.3\textwidth]{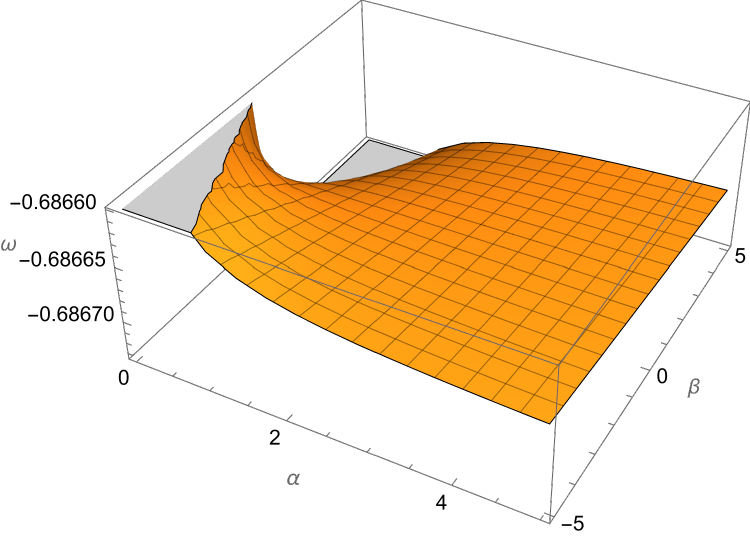}

\caption{This figure presents a comprehensive visualization of the energy conditions and the EoS parameter for Model 1, plotted as functions of the free parameters $\alpha$ and $\beta$.}
\label{F_ECs1}
\end{figure}

Fig. \ref{F_ECs1} shows the behavior of the energy conditions: WEC, NEC, DEC, and SEC for Model 1 as functions of the model parameters $\alpha$ and $\beta$. Below, we analyze the trends and implications of each energy condition: 

The WEC ensures that the energy density $\rho \geq 0$. to ensure $\rho > 0$, the following condition must hold: $6 \alpha H_0^2+\beta > 0$. For $\alpha > 0$: since $6 \alpha H_0^2 > 0$, the term $\beta$ must satisfy: $\beta > -6 \alpha H_0^2$. For $\alpha < 0$: since $6 \alpha H_0^2 < 0$, the term $\beta$ must satisfy:  $\beta < -6 \alpha H_0^2$. From Fig. \ref{F_ECs1} (first graph), $\rho$ increases linearly with both $\alpha$ and $\beta$. This linear dependence on the parameters aligns with the analytical expression in Eq. (\ref{WEC1}). Physically, this suggests that for sufficiently large values of $\alpha$ or $\beta$, the energy density remains positive, satisfying the WEC. The NEC is represented by Eq. (\ref{NEC1}). Fig. \ref{F_ECs1} (second graph) shows that NEC depends primarily on $\alpha$ and is relatively insensitive to $\beta$, consistent with the analytical form in Eq. (\ref{NEC1}). Positive values of $q_0 + 1$ amplify $\rho+p$, indicating the NEC is satisfied for certain ranges of $\alpha$. The DEC imposes the condition in Eq. (\ref{DEC1}). Fig. \ref{F_ECs1} (third graph) shows that $\rho-p$ has a linear dependence on both $\alpha$ and $\beta$. The behavior indicates that DEC is satisfied in regions where $\beta$ dominates over the second term. This implies that the choice of $\beta$ is critical in ensuring physical viability.

The SEC, defined by $\rho+3p \geq 0$, is heavily influenced by $\alpha$ and $\beta$. The figure reveals a strong dependence on $\beta$, with regions of negative values when $\beta$ becomes dominant. This highlights the interplay between $\alpha$ and $\beta$ in determining whether the SEC is satisfied. For high values of $\beta$, the SEC is likely violated. The violation of the SEC is crucial for explaining the accelerated expansion phase of the universe \cite{Visser/2000}. The SEC typically holds in a decelerating universe where gravity behaves attractively. However, the SEC must be violated during the acceleration phase (such as in the current epoch of dark energy domination or the early universe's inflationary phase). In Model 1, the SEC is given by Eq. (\ref{SEC1}), where the deceleration parameter $q_0$ plays a key role. For an accelerating universe ($q_0 < 0$), the term $6\alpha H_0^2 q_0$ becomes negative, promoting SEC violation. The parameter $\beta$, acting as an offset, determines the extent of the violation; larger $\beta$ values can counteract this effect and potentially satisfy the SEC unless carefully balanced. Observations from Fig. \ref{F_ECs1} (fourth graph) confirm that SEC violation ($\rho+3p < 0$) occurs in regions where $\beta$ and $q_0$ align to yield negative values, essential for cosmic acceleration. This behavior highlights how $f(Q, L_m)$ gravity allows for effective repulsive dynamics, countering standard gravitational attraction and enabling accelerated expansion. Proper tuning of $\alpha$, $\beta$, and $q_0$ is therefore critical for ensuring the model reproduces the observed accelerated universe.

For Model 1, the EoS parameter is
\begin{equation}
\label{w1}
\omega=-1+\frac{4 \alpha  H_0^2 (q_0+1)}{6 \alpha H_0^2+\beta}.  
\end{equation}

From Eq. (\ref{w1}), it is evident that the EoS parameter depends on $\alpha$ and $\beta$. The parameter $\alpha$ controls the strength of the $Q$ term in the model and directly scales the numerator. A larger $\alpha$ increases the contribution of $4 \alpha H_0^2 (q_0 + 1)$, affecting $\omega$. $\beta$ dominates the denominator for large values, suppressing the effect of the numerator and driving $\omega$ toward $-1$. The deceleration parameter $q_0$ determines the acceleration phase of the universe. For an accelerating universe ($q_0 < 0$), $\omega$ decreases, approaching or dropping below $-1$, mimicking dark or phantom energy behavior. $\omega = -1$ achieved when the denominator ($\beta + 6 \alpha H_0^2$) dominates over the numerator, indicating a cosmological constant-like behavior. $\omega > -1$ represents quintessence-like behavior, occurring when $4 \alpha H_0^2 (q_0 + 1)$ is positive but smaller than the denominator. $\omega < -1$ represents phantom energy, possible if $q_0 + 1$ becomes highly negative, enhancing the numerator \cite{Hernandez,Jesus,Cunha}. Fig. \ref{F_ECs1} (fifth graph) shows $\omega$ as a function of $\alpha$ and $\beta$, likely highlights these trends. Regions where $\omega \approx -1$ suggest the dominance of $\beta$, whereas deviations above or below $-1$ reflect the interplay between $\alpha$ and $\beta$. A sharp transition or peak in the plot indicates sensitivity to parameter variations.  

\subsection{Model 2: $f(Q,L_m)=-\alpha Q+ \lambda (2L_{m})^{2}+\beta$ with $L_{m}=\rho$}

As a second model, we consider $f(Q, L_m) = -\alpha Q + \lambda (2L_m)^2 + \beta$, where $\alpha$, $\beta$, and $\lambda$ are free parameters. This model, proposed by Hazarika et al. \cite{Hazarika/2024}, incorporates non-linear contributions from the matter Lagrangian. It is also capable of describing an accelerating universe. For this formulation, we obtain $f_Q = -\alpha$ and $f_{L_m} = 8 \lambda L_m$. Therefore, the Friedmann equations (\ref{F1}) and (\ref{F2}) reduce to
\begin{align}
\label{F1M2}
\rho &= \frac{\sqrt{6 \alpha  H^2+\beta}}{2 \sqrt{3\lambda}}, \\ 
p &= -\frac{6 \alpha  \left(\dot{H}+H^2\right)+\beta}{2 \sqrt{3 \lambda} \sqrt{6 \alpha  H^2+\beta}}.
\label{F2M2}
\end{align}

By substituting Eqs. (\ref{F1M2}), (\ref{F2M2}), and (\ref{dH}) into the energy conditions, we derive the following constraints:
\begin{equation} \label{WEC2}
WEC \Leftrightarrow \frac{\sqrt{6 \alpha H_0^2+\beta}}{2 \sqrt{3 \lambda}} \geq 0,
\end{equation}
\begin{equation} \label{NEC2}
NEC \Leftrightarrow \frac{\sqrt{3} \alpha H_0^2 (q_0 + 1)}{\sqrt{\lambda(6 \alpha H_0^2+\beta )}} \geq 0,
\end{equation}
\begin{equation} \label{DEC2}
DEC \Leftrightarrow \frac{\beta - 3 \alpha H_0^2 (q_0 - 1)}{\sqrt{3 \lambda} \sqrt{6 \alpha H_0^2+\beta}} \geq 0,
\end{equation}
\begin{equation}  \label{SEC2}
SEC \Leftrightarrow \frac{3 H_0^2 (\alpha + 3 \alpha q_0) - \beta}{\sqrt{3 \lambda} \sqrt{6 \alpha H_0^2+\beta}} \geq 0.
\end{equation}

\begin{figure}[H]
\centering
\includegraphics[width=.3\textwidth]{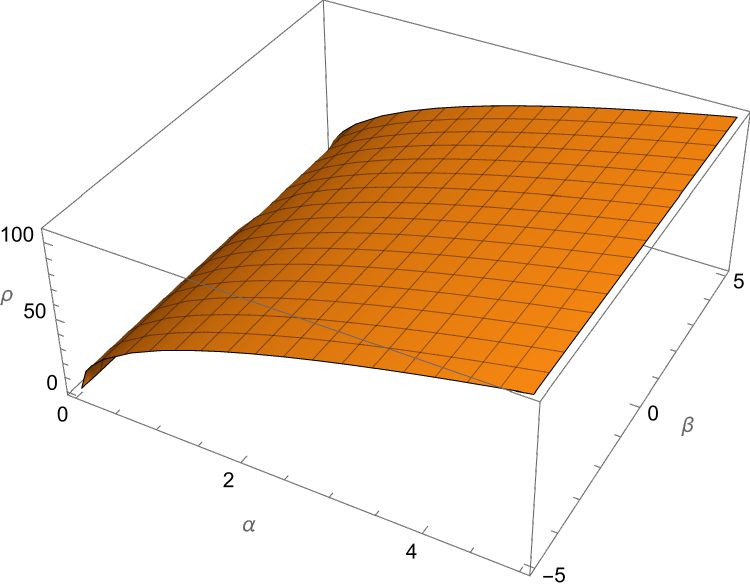}\quad
\includegraphics[width=.3\textwidth]{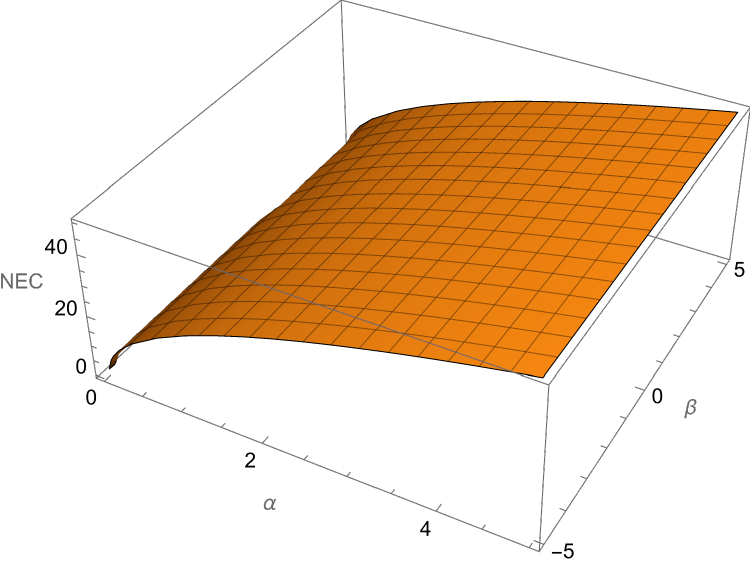}\quad
\includegraphics[width=.3\textwidth]{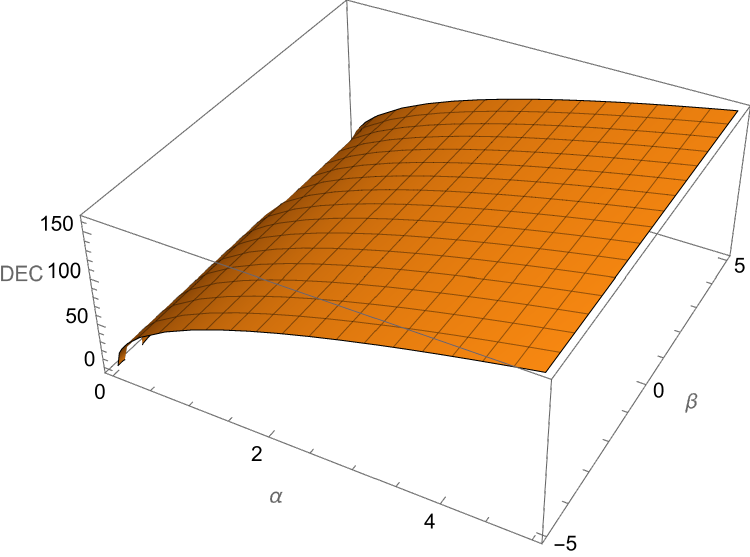}

\medskip

\includegraphics[width=.3\textwidth]{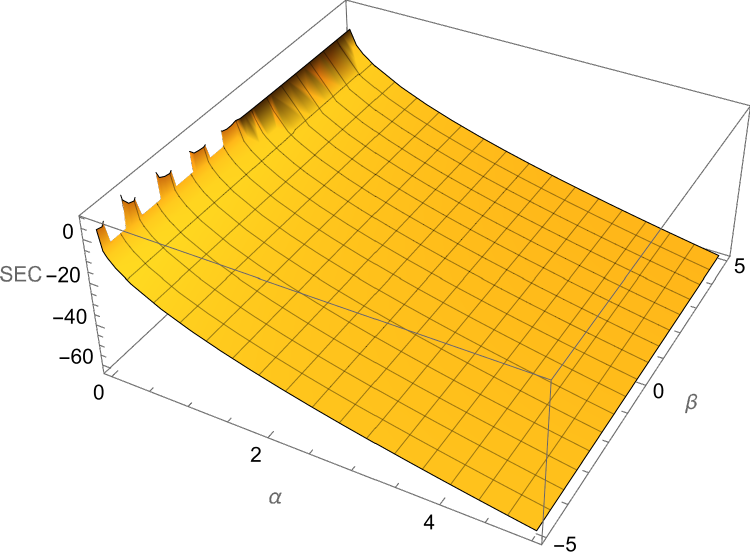}\quad
\includegraphics[width=.3\textwidth]{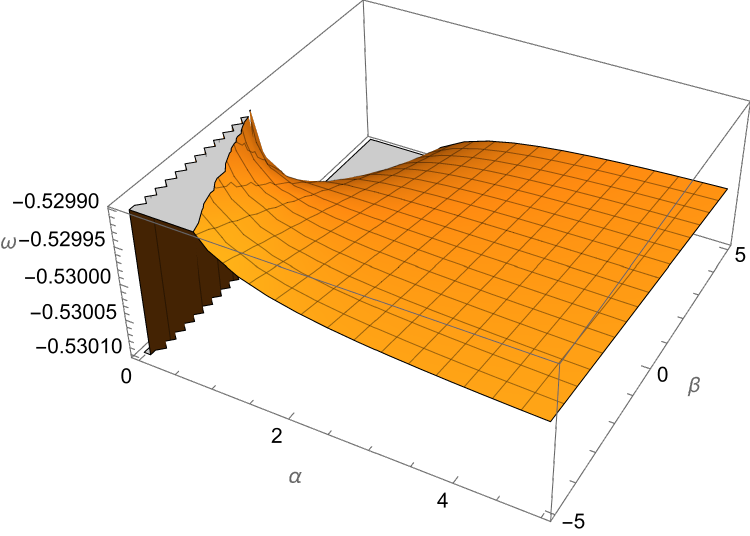}

\caption{This figure presents a comprehensive visualization of the energy conditions and the EoS parameter for Model 2, plotted as functions of the free parameters $\alpha$ and $\beta$.}
\label{F_ECs2}
\end{figure}

Fig. \ref{F_ECs2} shows the behavior of the energy conditions: WEC, NEC, DEC, and SEC for Model 2 as functions of $\alpha$ and $\beta$. To reduce the number of parameters, we consistently set $\lambda = 1$, since $\lambda$ is a scaling parameter. Below is a detailed interpretation of each plot:

The WEC ensures non-negativity of the energy density $\rho$, which for Model 2 is given by the expression in Eq. (\ref{WEC2}). Fig. \ref{F_ECs2} (first graph) shows that $\rho$ increases monotonically with both $\alpha$ and $\beta$. Larger values of $\beta$ dominate the expression, pushing $\rho$ higher, while $\alpha$ contributing through $6\alpha H_0^2$, enhancing $\rho$. This behavior aligns with physical expectations that the energy density remains positive for viable cosmological models. The NEC is sensitive to $q_0 + 1$, meaning it is satisfied when the universe transitions to or remains in an accelerating phase ($q_0 + 1 \geq 0$). Fig. \ref{F_ECs2} (second graph) indicates NEC is satisfied across the range of $\alpha$ and $\beta$, with $\rho+p$ increasing as $\alpha$ or $\beta$ rises. Fig. \ref{F_ECs2} (third graph) shows $\rho-p$ is positive for a wide range of $\alpha$ and $\beta$, indicating that the DEC is generally satisfied. For larger $\beta$, $\rho$ dominates over $p$, stabilizing the condition. The interplay between $\beta$ and $q_0$ ensures the DEC aligns with physical expectations for most parameter values.  

In addition, Fig. \ref{F_ECs2} (fourth graph) indicates that SEC is violated ($\rho+3p < 0$) in regions where $\beta$ dominates, consistent with the need to violate the SEC to explain the universe's accelerated expansion. The degree of violation increases for negative $q_0$ (accelerating universe) and higher $\beta$, demonstrating the model's capability to describe dark energy-like behavior.  

For Model 2, the EoS parameter is
\begin{equation}
\omega=\frac{6 \alpha  H_0^2 q_0-\beta }{6 \alpha  H_0^2+\beta}.  
\end{equation}

Fig. \ref{F_ECs2} (fifth graph) shows that $\omega$ depends on the balance between $6\alpha H_0^2 q_0$ and $\beta$. For higher $\beta$, $\omega$ approaches $-1$, resembling a cosmological constant-like behavior. For smaller $\beta$ or more negative $q_0$, $\omega$ drops further, potentially entering the phantom energy regime ($\omega < -1$). The small variations in the figure reflect the sensitivity of $\omega$ to changes in $\alpha$, $\beta$, and $q_0$.

\subsection{Summary of key differences}

The two models of $f(Q, L_m)$ gravity, Model 1 ($f(Q, L_m) = -\alpha Q + 2L_m + \beta$) and Model 2 ($f(Q, L_m) = -\alpha Q + \lambda (2L_m)^2 + \beta$), show distinct behaviors for the energy conditions and the EoS. Below is a detailed comparison (please see Tab. \ref{tab}):

\begin{table}[H]
    \centering
    \begin{tabular}{|c|c|c|}
    \hline
    \textbf{Aspect} & \textbf{Model 1} & \textbf{Model 2} \\ \hline
    \text{WEC} (\(\rho\)) & \text{Linear dependence on } \(\alpha, \beta\) & \text{Nonlinear dependence (square root form)} \\ \hline
    \text{NEC} (\(\rho + p\)) & \text{Proportional to } \(q_0 + 1\) & \text{Modulated by } \(\sqrt{6\alpha H_0^2+\beta}\) \\ \hline
    \text{DEC} (\(\rho - p\)) & \text{Linear in } \(\beta\) & \text{Nonlinear, with wider parameter satisfaction} \\ \hline
    \text{SEC} (\(\rho + 3p\)) & \text{Limited violation for acceleration} & \text{Significant violation for acceleration} \\ \hline
    \text{EoS} (\(\omega\)) & \text{Simpler, approaches } -1 \text{ for large } \(\beta\) & \text{Flexible, capable of phantom and quintessence} \\ \hline
    \end{tabular}
    \caption{Comparative analysis of energy conditions and EoS in two models of $f(Q, L_m)$ gravity.}
    \label{tab}
\end{table}
        
In conclusion, Model 1 offers simplicity and is suitable for basic cosmological analysis, particularly in linear regimes. In contrast, Model 2 is more robust, flexible, and capable of capturing complex dynamics, such as SEC violations and transitions between quintessence and phantom energy. Thus, Model 2 is better suited for explaining accelerated cosmic expansion and dark energy phenomena.

\section{Conclusions}\label{sec5}

Energy conditions play a pivotal role in establishing a consistent and physically viable theory of gravity. As novel theories of gravity continue to emerge in the literature, it becomes increasingly important to evaluate them through the lens of energy condition constraints \cite{Capozziello/2018,Atazadeh/2009,Liu/2012,Zubair/2015,Azizi/2017,Garcia/2011,Bamba/2017,Sharif/2016,Yousaf/2018,Mandal/2020,Arora/2021}. In this study, we analyzed the weak, null, dominant, and strong energy conditions for two $f(Q, L_m)$ gravity models. The $f(Q, L_m)$ framework, which combines the non-metricity scalar $Q$ with the matter Lagrangian $L_m$, represents a promising extension of gravitational theory \cite{Hazarika/2024}.

To analyze the energy conditions, we used the homogeneous and isotropic FLRW metric, which describes cosmological evolution within a flat geometry. In this work, we explored two specific classes of cosmological models by considering simple functional forms of $f(Q, L_m)$. For the first model, we adopted an additive Lagrangian given by $f(Q, L_m) = -\alpha Q + 2L_m + \beta$. This Model 1 exhibits linear dependence on the parameters $\alpha$ and $\beta$, making it analytically simpler and more tractable for basic cosmological analysis. The WEC, NEC, and DEC are generally satisfied for appropriate ranges of $\alpha$ and $\beta$, ensuring the model's physical viability. The SEC, however, is selectively violated, a feature essential for describing the accelerated expansion of the universe. This SEC violation occurs in regions where $\beta$ and the deceleration parameter $q_0$ balance to yield negative values for $\rho + 3p$, aligning with observations of dark energy-driven cosmic acceleration. The EoS parameter $\omega$ transitions between quintessence-like behavior ($\omega > -1$), a cosmological constant ($\omega \approx -1$), and phantom energy ($\omega < -1$), depending on the interplay of $\alpha$, $\beta$, and $q_0$. This adaptability makes Model 1 a viable candidate for studying cosmic acceleration, albeit with limitations in capturing nonlinearities. The second model is characterized by the functional form $f(Q, L_m) = -\alpha Q + \lambda (2L_m)^2 + \beta$, introduces nonlinearities, offering greater flexibility and robustness in describing complex cosmological dynamics. The WEC, NEC, and DEC are satisfied across a broader range of parameters, indicating strong physical consistency. The SEC is consistently violated in regions dominated by $\beta$ and negative $q_0$, reinforcing the model's suitability for explaining dark energy and the universe's accelerated expansion. The nonlinear dependence of $\rho$, $p$, and $\omega$ on $\alpha$ and $\beta$ allows for a richer variety of behaviors, including transitions between quintessence and phantom energy. The sensitivity of $\omega$ to parameter changes highlights the model's capacity to capture subtle variations in the universe's expansion dynamics.

Thus, Model 1 offers simplicity, making it ideal for linear cosmological scenarios and analytical explorations. In contrast, Model 2's nonlinear structure enables it to address complex phenomena like stronger SEC violations and a broader parameter space for accelerated expansion. These features make Model 2 better suited for describing the universe's current and early acceleration phases, particularly in scenarios involving dark energy transitions. The $f(Q, L_m)$ gravity framework demonstrates its potential as an extension of standard cosmology. By appropriately tuning the parameters $\alpha$ and $\beta$, both models can reproduce observed cosmic behaviors. A crucial next step is to test the identified parameter space against observational data, including supernovae type Ia (SNe Ia), cosmic microwave background (CMB), baryon acoustic oscillations (BAO), and large-scale structure (LSS). Future work will focus on statistical analyses, such as Bayesian inference and Markov Chain Monte Carlo (MCMC) methods, to determine best-fit parameter values and assess the observational viability of these $f(Q, L_m)$ models.

\section*{Acknowledgment}
This research was funded by the Science Committee of the Ministry of Science and Higher Education of the Republic of Kazakhstan (Grant No. AP22682760).

\section*{Data Availability Statement}
This article does not introduce any new data.


\begin{thebibliography}{99}

\bibitem{Riess/1998} A.G. Riess et al., \textit{Astron. J.} \textbf{116} 1009 (1998).

\bibitem{Riess/2004} A.G. Riess et al., \textit{Astophys. J.} \textbf{607} 665-687 (2004).

\bibitem{Perlmutter/1999} S. Perlmutter et al., \textit{Astrophys. J.} \textbf{517} 377 (1999).

\bibitem{T.Koivisto} T. Koivisto, D.F. Mota, \textit{Phys. Rev. D} \textbf{73}, 083502 (2006).

\bibitem{S.F.} S.F. Daniel, \textit{Phys. Rev. D} \textbf{77}, 103513 (2008).

\bibitem{Spergel} D.N. Spergel et al., \textit{Astrophys.
J. Suppl.} \textbf{148}, 175 (2003).

\bibitem{R.R.} R.R. Caldwell, M. Doran, \textit{Phys. Rev. D} \textbf{69}, 103517 (2004).

\bibitem{Z.Y.} Z.Y. Huang et al., \textit{J. Cosm. Astrop. Phys.} \textbf{0605}, 013 (2006).

\bibitem{D.J.} D.J. Eisenstein et al., \textit{Astrophys. J.} \textbf{633}, 560 (2005).

\bibitem{W.J.} W.J. Percival at el., \textit{Mon. Not. R. Astron. Soc.} \textbf{401}, 2148 (2010).

\bibitem{Zlatev/1999} I. Zlatev, L. Wang, and P.J. Steinhardt, \textit{Phys. Rev. Lett.} \textbf{82} 896 (1999).

\bibitem{Weinberg/1989} S.Weinberg,  \textit{Rev. Mod. Phys.} \textbf{61} 1 (1989).

\bibitem{Padmanabhan/2003} T. Padmanabhan,  \textit{Phys. Rep.} \textbf{380} 235 (2003).

\bibitem{Steinhardt/1999} P.J. Steinhardt, L. Wang, and I. Zlatev, \textit{Phys. Rev. D} \textbf{59} 123504 (1999).

\bibitem{Buchdahl/1970} H.A. Buchdahl, \textit{Mon. Not. R. Astron. Soc.} \textbf{150} 1 (1970).

\bibitem{Dunsby/2010} P.K.S. Dunsby et al., \textit{Phys. Rev. D} \textbf{82} 023519 (2010).

\bibitem{Carroll/2004} S.M. Carroll et al., \textit{Phys. Rev. D} \textbf{70} 043528 (2004).

\bibitem{Harko/2010} T. Harko and F.S.N. Lobo, \textit{Eur. Phys. J. C} \textbf{70} 373–379 (2010).

\bibitem{Wang/2012} J. Wang and K. Liao, \textit{Class. Quantum Gravity} \textbf{29} 215016 (2012).

\bibitem{Goncalves/2023} B.S. Goncalves and P.H.R.S. Moraes, \textit{Fortschr. Phys.} \textbf{71} 2200153 (2023).

\bibitem{Myrzakulova/2024} S. Myrzakulova et al., \textit{Phys. Dark Universe} \textbf{43} 101399 (2024).

\bibitem{Myrzakulov/2024} Y. Myrzakulov et al., \textit{Phys. Dark Universe} \textbf{45} 101545 (2024).

\bibitem{Felice/2009} A. De Felice and S. Tsujikawa, \textit{Phys. Lett. B} \textbf{675} 1 (2009).

\bibitem{Bamba/2017} K. Bamba et al., \textit{Gen. Relativ. Gravit.} \textbf{49} 112 (2017).

\bibitem{Goheer/2009} N. Goheer et al., \textit{Phys. Rev. D} \textbf{79} 121301 (2009).

\bibitem{Harko/2011} T. Harko et al., \textit{Phys. Rev. D} \textbf{84} 024020 (2011).

\bibitem{Koussour_1/2022} M. Koussour and M. Bennai, \textit{Int. J. Geom. Methods Mod. Phys.} \textbf{19} 2250038 (2022).

\bibitem{Koussour_2/2022} M. Koussour and M. Bennai, \textit{Afr. Mat.} \textbf{33} 27 (2022).

\bibitem{Myrzakulov/2023} N. Myrzakulov et al., \textit{Chin. Phys. C} \textbf{47} 115107 (2023).

\bibitem{KK1} M. Koussour et al., \textit{Phys. Dark Universe} \textbf{46} 101577 (2024).

\bibitem{Jimenez/2018} J.B. Jimenez, L. Heisenberg, and T. Koivisto, \textit{Phys. Rev. D} \textbf{98} 044048 (2018).

\bibitem{Jimenez/2020} J.B. Jimenez et al., \textit{Phys. Rev. D} \textbf{101} 103507 (2020).

\bibitem{Khyllep/2021} W. Khyllep et al., \textit{Phys. Rev. D} \textbf{103} 103521 (2021).

\bibitem{MK1} M. Koussour et al., \textit{Phys. Dark universe} \textbf{36},
101051 (2022).

\bibitem{MK2} M. Koussour and M. Bennai \textit{Chin. J. Phys. } \textbf{379},
339-347 (2022).

\bibitem{MK3} M. Koussour et al. \textit{Phys. Ann. Phys.} \textbf{445},
169092 (2022).

\bibitem{MK4} M. Koussour and A. De, \textit{Eur. Phys. J. C} \textbf{83}, 400 (2023).

\bibitem{MK5} M. Koussour et al., \textit{Fortschr. Phys.} \textbf{71}, 2200172 (2023).

\bibitem{MK6} M. Koussour et al., \textit{Nucl. Phys. B} \textbf{990}, 116158 (2023).

\bibitem{MK7} M. Koussour et al., \textit{J. High Energy Phys.} \textbf{37}, 15-24 (2023).

\bibitem{MK8} M. Koussour et al., \textit{J. High Energy Astrophys, } \textbf{35}, 43-51 (2022).

\bibitem{Xu/2019} Y. Xu et al., \textit{Eur. Phys. J. C} \textbf{79}, 708 (2019).

\bibitem{Xu/2020} Y. Xu et al., \textit{Eur. Phys. J. C} \textbf{80}, 449 (2020).

\bibitem{K6}  M. Koussour et al., \textit{Int. J. Mod. Phys. D} \textbf{31},  2250115 (2022).

\bibitem{K7}  M. Koussour et al., \textit{Chin. J. Phys.} \textbf{86},  300-312 (2023).

\bibitem{Bourakadi} K. El Bourakadi et al., \textit{Phys. Dark Universe} \textbf{41}, 101246 (2023).

\bibitem{KK2} M. Koussour et al., \textit{Phys. Dark Universe} \textbf{45}, 101527 (2024).

\bibitem{KK3} M. Koussour et al., \textit{Chin. J. Phys.} \textbf{90}, 108-120 (2024).

\bibitem{Hazarika/2024} A. Hazarika et al. \textit{arXiv}, arXiv:2407.00989 (2024).

\bibitem{fRLT} Z. Haghani and T. Harko, \textit{Eur. Phys. J. C} \textbf{81}, 615 (2021).

\bibitem{Mustafa} G. Mustafa et al., \textit{Phys. Dark Universe} \textbf{45}, 101508 (2024).

\bibitem{fQL1} Y. Myrzakulov et al., \textit{Phys. Dark Universe} \textbf{46}, 101614 (2024).

\bibitem{fQL2} K. Myrzakulov et al., \textit{J. High Energy Astrophys.} \textbf{44}, 164-171 (2024).

\bibitem{fQL3} Y. Myrzakulov et al., \textit{Phys. Dark Universe} \textbf{48}, 101829 (2025).

\bibitem{Visser/2000} M. Visser and C. Barcelo, \textit{Cosmo-99}, 98-112 (2000).

\bibitem{Hawking/1973} S.W. Hawking and G.F.R. Ellis, \textit{The Large Scale Structure of Space-Time, Cambridge University Press}, 98-112 (1973).

\bibitem{Wald/1984} R.M. Wald, \textit{General relativity, University of Chicago Press}, (1984).

\bibitem{Santos/2007} J. Santos et al., \textit{Phys. Rev. D} \textbf{76}, 083513 (2007).

\bibitem{Visser/1996} M. Visser, \textit{Lorentzian wormholes. from Einstein to Hawking. Woodbury} (1996).

\bibitem{Capozziello/2018} S. Capozziello, S. Nojiri, and S.D. Odinstov, \textit{Phys. Lett B} \textbf{781}, 99 (2018).

\bibitem{Atazadeh/2009} K. Atazadeh, A. Khaleghi, H.R. Sepangi, and Y. Tavakoli, \textit{Int. J. Mod. Phys. D} \textbf{18}, 1101 (2009).

\bibitem{Liu/2012} D. Liu and M.J. Reboucas, \textit{Phys. Rev. D} \textbf{86}, 083515 (2012).

\bibitem{Zubair/2015} M. Zubair and S. Waheed, \textit{Astrophys. Space Sci.} \textbf{355}, 361 (2015).

\bibitem{Azizi/2017} T. Azizi and M. Gorjizadeh, \textit{Europhys. Lett.} \textbf{117}, 60003 (2017).

\bibitem{Garcia/2011} N.M. Garcia et al., \textit{Phys. Rev. D} \textbf{83}, 104032 (2011).

\bibitem{Bamba/2017} K. Bamba et al., \textit{Gen. Relativ. Gravit.} \textbf{49}, 112 (2017).

\bibitem{Sharif/2016} M. Sharif and A. Ikram, \textit{Eur. Phys. J. C} \textbf{76}, 640 (2016).

\bibitem{Yousaf/2018} Z. Yousaf et al., \textit{Int. J. Geom. Methods Mod. Phys.} \textbf{15}, 1850146 (2018).

\bibitem{Mandal/2020} S. Mandal, P.K. Sahoo, and J.R.L. Santos, \textit{Phys. Rev. D} \textbf{102}, 024057 (2020).

\bibitem{Arora/2021} S. Arora, J.R.L. Santos, and P.K. Sahoo, \textit{Phys. Dark Universe	} \textbf{31}, 100790 (2021).

\bibitem{ryden/2003} B. Ryden, \textit{ Introduction to Cosmology} (Addison Wesley, San Francisco, United States of America, 2003).

\bibitem{Iosifidis/2018} D. Iosifidis, C.G. Tsagas, and A.C. Petkou, \textit{Phy. Rev. D} \textbf{98}, 104037 (2018).

\bibitem{Yang/2021} J.Z. Yang et al., \textit{Eur. Phys. J. C} \textbf{81}, 1-19 (2021).

\bibitem{Planck/2014} Planck Collaboration XVI, \textit{Astron. Astrophys.}, \textbf{571}, A16 (2014). 

\bibitem{Planck/2020} N. Aghanim et al. (Planck), \textit{Astron. Astrophys.}, \textbf{641}, A6 (2020).

\bibitem{w1} Y. Myrzakulov et al.,  \textit{J. High Energy Astro. Phys.}, \textbf{43}, 209-216 (2024).

\bibitem{w2} A. Errehymy et al.,  \textit{Phys. Dark Universe	}, \textbf{46}, 101555 (2024).

\bibitem{Hernandez} A. Hernandez-Almada et al., \textit{Eur. Phys. J. C}, \textbf{79}, 12 (2019).

\bibitem{Jesus} J.F. Jesus et al., \textit{J. Cosmol. Astropart. Phys.}, \textbf{04}, 053 (2020).

\bibitem{Cunha} J.V. Cunha and J.A.S. Lima, \textit{Mon. Not. R. Astron. Soc.}, \textbf{390}, 210 (2008).

\end{thebibliography}
\end{document}